\documentclass{PoS}
\usepackage{float}

\title{Experimentally, How Dark Are Black Hole Mergers?}

\ShortTitle{DESGW}

\author{\speaker{James Annis}
        Fermilab\\
        \email{annis@fnal.gov}}
\author{Marcelle Soares-Santos
        Fermilab\\
        \email{marcelle@fnal.gov}}

\abstract {
    The first Advanced LIGO observing run detected two black hole merger events
    with confidence and likely a third. Many groups organized to
    followup the events in the optical even though the strong theoretical prior that
    no optical emission should be seen. We carry through the logic of this
    by asking about the experimental upper limits to the optical light from 
    Advanced LIGO black hole mergere events. We inventory the published
    optical searches for transient events associated with the black hole
    mergers. We describe the factors that go into a formal limit on
    the visibility of an event (sky area coverage, the coverage factor of the
    camera, the fraction of sky not covered by intervening objects),
    and list what is known from the literature
    of the followup teams quantitative assessment of each factor. Where
    possible we calculate the total probability from each group that the source was imaged.
    The calculation of confidence level is reviewed for the case of no background. We 
    find that an experimental 95\% upper limit on the magnitude of a black hole
    requires the sum of the total probabilities over all events to be more than 3.
    In the first Advanced LIGO observing run we were far from reaching that threshold.
          }

\FullConference{38th International Conference on High Energy Physics\\
		3-10 August 2016\\
		Chicago, USA}

\begin{document}

\section{Are black hole mergers dark?}

There is a strong theoretical prior that the black hole mergers
detected by Advanced LIGO \cite{black_holes} were invisible in the optical.
The expected progenitor systems lack
massive accretion disks to feed into the jets that make gamma-ray bursts and AGN so visible.
Solar-mass scale accretion disks are, however, at least 
a logical possibility (e.g., \cite{perna2016}, \cite{murase2016}) 
and even without them black holes in astrophysical environments
are merely dim, not invisible (e.g., \cite{heckler1996, chisholm2003, beskin2005}).
Furthermore, there are exotic compact objects that can have the redshift-spheres  of black holes
but which may or may not posses an event horizon (e.g., boson stars, dark matter stars, dark energy stars)
see \cite{giudice2016}); whether or not these have optical signatures different than 
black holes is a research topic. We are led to ask the question:
experimentally, how certain are we that there is no optical signature 
from the Advanced LIGO black hole merger events? 

\section{The event inventory and optical searches}

The LIGO Science Collaboration discovered 3 black hole mergers in their
first observing run; see table 1 (data from \cite{black_holes}).
They confidently detected two merging systems, and likely detected a third.
Only confidently detected events were transmitted to the electromagnetic followup teams.

\begin{table}[H]
\begin{tabular}{r r r r r c c}
               & distance & final mass & $1^{st}$ mass & $2^{nd}$  mass& 90\% sky area & significance \\
               & Mpc      &   M$_\odot$     &   M$_\odot$ &  M$_\odot$       &   deg$^2$   &    $\sigma$ \\
GW150914  & 420$^{+150}_{-180}$ & 62.3$^{+3.7}_{-3.1}$ & 36.2$^{+5.2}_{-3.8}$ & 29.1$^{+3.7}_{-4.4}$ & 230 & $>5.3$ \\
GW151226 & 440$^{+180}_{-190}$ & 20.8$^{+6.1}_{-1.7}$ & 14.2$^{+8.3}_{-3.7}$ & 7.5$^{+2.3}_{-2.3}$ & 850 & $>5.3$ \\
LVT 151012 & 1000$^{+500}_{-500}$ & 35$^{+14}_{-4}$ & 23$^{+18}_{-6}$ & 13$^{+4}_{-5}$ & 1600 & $1.7$ \\
\end{tabular}
\caption{Measurements of the three most significant events:
median values with 90\% credible intervals. 
}
\end{table}

%

As of this writing (September 2016) there have been 10 papers describing followup in the optical.
More reports of observations have been sent to GCN (see the list in \cite{emligo2016}); 
we will put these aside, as we will the non-optical followup.

The optical followup breaks into two deep wide-field searches,
two shallow very wide-field searches, and two searches based on nearby galaxies,
The deep wide-field searches are from the DESGW and Pan-STARRS.
Our DESGW group produced two papers on GW150914
\cite{marcelle2016, annis2016}and one on GW151226 \cite{phil2016};
two describing the search for the events and one describing a search for failed supernovae.
The group using Pan-STARRS
produced two papers, the first  \cite{ps2016a}  describing the search for GW150914 in detail, the
second briefly describing the search for GW151226  \cite{ps2016b} (and mentioning ATLAS telescope data); 
these papers also described 
extensive transient followup observations (i.e, spectroscopy and further photometry) of candidates.
The shallow very wide field searches were by the 
collaboration operating the MASTER Global Robotic Network of 6 telescopes
\cite{master2016}, and  by the Japanese collaboration for gravitational wave electromagnetic followup
(J-GEM) which uses 17 telescopes \cite{jgem2016}.
The groups using nearby galaxies to choose where to point their telescopes were
TOROS collaboration \cite{toros2016} and the group using the iPTF \cite{iptf2016}.

%
%


\section{Limits on optical signatures from black hole mergers}

What is the probability of detecting a light source above a certain magnitude?
Start with 
the probability $P$ that an imaging element was able to measure a source:
\begin{equation}
   P = \Sigma_{spatial} \cdot \epsilon_{camera} \cdot \epsilon_{area}\, .
\end{equation}
Table 2 tabulates the information from the literature, where:
\begin{itemize}
\item
$\Sigma_{spatial}$ is the summed probability inside the Advanced LIGO spatial localization map
covered by the bounding box of images taken.
\item $\epsilon_{camera}$ is the fraction of the camera that is live: the DECam, for example has an imaging
area of $\approx \pi$ deg$^2$, but only 80\% of it is filled with useful silicon; the rest
are gaps between the CCDs, dead CCDs, or glowing edge regions around the perimeter of the CCDs
that we remove from the analysis. 
DESGW and iPTF reported this number; PanSTARRS did not, perhaps because
they covered the area multiple times (which has the effect of $\epsilon_{camera} \rightarrow 1$).
\item $\epsilon_{sky}$ is the fraction of the area imaged that a source would have been
visible in. The DESGW, iPTF, and PanSTARRS analysis used fake objects injected into the images
and recovered to measure this. Unfortunately, both iPTF and PanSTARRS used the measurement
to establish limiting magnitude by locating the magnitude at which 50\% of the fakes were found;
it would be better to separate the two ideas.
\end{itemize}

\begin{table}
\begin{tabular}{l l c c c c c}
Event & Experiment & Magnitude 
    & \boldmath${\Sigma_{spatial}}$ & \boldmath${\epsilon_{camera}}$ & 
    \boldmath${\epsilon_{sky}}$ & \boldmath${P_{total}}$ \\
\\
GW150914 & Master 2x0.4m         & 18.4-19.9 O & 49\%   & 1.0          & \\
    & DESGW 4.0m Blanco, DECam   & 22.1 i      & 11\%   & 0.8 $\cdot$ 0.34\textsuperscript{ \textdagger} 
                                                                       & 0.8     & .024 \\
    & PanSTARRS 1.8m             & 19.7 i      & 4.2\%  & 1.0\textsuperscript{ \textdagger\textdagger} 
                                                                        & 0.5\textsuperscript{ \textdaggerdbl} 
                                                                                 & .021 \\
    & iPTF 1.1m Oschint, CFH12K  & 20.6 R      & 0.2\%  & 1.0\textsuperscript{ \textsection}
                                                                       &  & \\
    & JGEM 1.1m Kiso,  KWFC      & 19.0 i      & 0.1\%  &              &  \\
    & TOROS 1.5m EABA            & 21.7 r      & galaxy & 1.0 \\
\\
GW151226 & ATLAS 0.5m             & 19.0 o      & 36\% \\
         & PanSTARRS 1.8m         & 20.5 i      & 26.5\% &  1.0\textsuperscript{ \textdagger\textdagger} 
                                                                        & 0.5\textsuperscript{ \textdaggerdbl} 
                                                                                 & .133 \\
    & DESGW 4.0m Blanco,  DECam   & 22.1 i      & 2\%    & 0.8          & 0.8     & .013\\
\end{tabular}
\caption{Published optical followup data on Advanced LIGO black hole mergers, showing
the basic information needed to place an upper limit.
The probability that we observed the location of the black hole merger is $P_{total}$. \newline
\textsuperscript{\textdagger} Template coverage.\newline
\textsuperscript{\textdagger\textdagger} Assuming the dither pattern covers the entire area.\newline
\textsuperscript{\textdaggerdbl} Sky fraction is mixed with, and dominated by, limiting magnitude.\newline
\textsuperscript{\textsection} The iPTF reported only the area after taking into account $\epsilon_{camera}$.
}
\end{table}

Next ask whether we imaged the precise sky location of at least one merger:
the probability that we would cover at least one merger is 
1 minus the probability of missing all events:
\begin{equation}
P_{one} = 1-\prod_i (1-P_i)
\end{equation}
where $P_i$ denotes $P_{total}$ for the $i^{th}$ event.
After 6 events, assuming $P_{total}$ = 50\% for all, $P_{one} = 0.984$; 
after 10, $P_{one} = 0.999$

Finally, ask about the upper limit one can place. Assuming no background and a non-zero
uniform prior,
the cumulative posterior PDF is:
\begin{equation}
F(s|0) = \int_0^s \frac{t^ne^{-t}}{n!} {\rm d}t = 1-e^{-s}
\end{equation}
where $s$ is the $\sum P_{tot}$ (accounting for sky overlaps; please let's put Healpix maps into GraceDB!).
If one wants to place a 95\% confidence limit at a given magnitude, one needs $s \ge 3$.
The result of optical followup of the first Advanced LIGO as presented in Table 2 has $s = 0.19$.
This could be raised by a factor of 2-3 by determining the $\epsilon_{sky}$ for 
the MASTER and ATLAS data and completing the template set for DESGW.

Experimentally we have no constraint on
the optical emission from a black hole merger. A reasonable confidence level upper limit should
be one aim of the optical followup to Advanced LIGO observing run 2.



\begin{thebibliography}{17}

\bibitem[1]{black_holes}
{LSC} {et~al.} 2016, arXiv:1606.04856

\bibitem[2]{perna2016} Perna, R., Lazzati, D., \& Giacomazzo, B.\ 2016, ApJL, 821, L18 [arXiv/1602.05140]

\bibitem[3]{murase2016} Murase, K., Kashiyama, K., M{\'e}sz{\'a}ros, P., et al.\ 2016, ApJL, 822, L9 [arXiv/1602.06938]

\bibitem[4]{heckler1996}
{Heckler}, A.~F., \& {Kolb}, E.~W. 1996, ApJL, 472, L85 [astro-ph/9605199]

\bibitem[5]{chisholm2003}
{Chisholm}, J.~R., {Dodelson}, S., \& {Kolb}, E.~W. 2003, ApJ, 596, 437 [astro-ph/0205138]

\bibitem[6]{beskin2005} Beskin, G.~M., \& Karpov, S.~V.\ 2005, Astronomy \& Astrophysics, 440 [astro-ph/0403649]

\bibitem[7]{giudice2016} Giudice, G.~F., McCullough, M., \& Urbano, A.\ 2016, arXiv:1605.01209 

\bibitem[8]{emligo2016} Abbott, B.~P., Abbott, R., Abbott, T.~D., et al.\ 2016, ApJL, 826, L13 [arXiv/1602.08492]

\bibitem[9] {marcelle2016}
    Soares-Santos, M., Kessler, R., Berger, E.,  et~al. 2016, ApJL, 823, L33 [arXiv/1602.04198]

\bibitem[10] {annis2016}
    Annis, J., Soares-Santos, M., Berger, E., {et~al.} 2016, ApJL, 823, L34 [arXiv/1602.04199]

\bibitem[11]{phil2016}
    Cowperthwaite, P.~S. et al. 2016, arXiv:1606.04538

\bibitem[12]{ps2016a} Smartt, S.~J., Chambers, K.~C., Smith, K.~W., et al.\ 2016, arXiv:1602.04156 

\bibitem[13]{ps2016b} Smartt, S.~J., Chambers, K.~C., Smith, K.~W., et al.\ 2016, arXiv:1606.04795 

\bibitem[14]{master2016} 
    Lipunov, V.~M., Kornilov, V., Gorbovskoy, E., et al.\ 2016, arXiv:1605.01607

\bibitem[15]{jgem2016} 
    Morokuma, T. et al.\ 2016, Pub. of the Astronomical Society of Japan, 68, L9 [arXiv/1605.03216]

\bibitem[16]{toros2016} 
    D{\'{\i}}az, M.~C., Beroiz, M., Pe{\~n}uela, T., et al.\ 2016, arXiv:1607.07850

\bibitem[17]{iptf2016}
     Kasliwal, M.~M., Cenko, S.~B., Singer, L.~P., et al.\ 2016, ApJL, 824, L24 [arXiv/1602.08764]

\end{thebibliography}
\end{document}